\documentclass[iop,apj]{emulateapj}
\slugcomment{The Astrophysical Journal, 833:224 (10pp), 2016 December 20} 
\shorttitle{Star Formation in Intermediate Redshift $0.2 < z < 0.7$  Brightest Cluster Galaxies}
\shortauthors{Cooke et al. 2016}

\usepackage{natbib}
\usepackage{graphicx} 
\usepackage{verbatim}
\usepackage[outdir=./]{epstopdf}
\usepackage{indentfirst}

\usepackage{amssymb}
\usepackage{amsmath}
\usepackage{fancyhdr}
\usepackage[hyperindex,breaklinks]{hyperref}

\begin{document}

\title{Star Formation in Intermediate Redshift $0.2 < z < 0.7$  Brightest Cluster Galaxies}

\author{Kevin C. Cooke$^1$, Christopher P. O'Dea$^{1,2}$, Stefi A. Baum$^{1,2}$, Grant R. Tremblay$^3$, Isabella G. Cox$^1$, and Michael D. Gladders$^4$}
\affil{$^1$School of Physics and Astronomy, Rochester Institute of Technology, Rochester, NY 14623, USA; \href{mailto:kcc7952@rit.edu}{kcc7952@rit.edu}\\
$^2$Department of Physics and Astronomy, University of Manitoba, Winnipeg, MB R3T 2N2, Canada\\
$^3$Department of Astronomy and Physics, Yale University, New Haven, CT 06511, USA\\
$^4$Department of Astronomy and Astrophysics and Kavli Institute for Cosmological Physics,\\ University of Chicago, 5640 South Ellis Avenue, Chicago, IL 60637, USA}


\begin{abstract}
We present a multi-wavelength photometric and spectroscopic study of 42 Brightest Cluster Galaxies (BCGs) in two samples of galaxy clusters chosen for a gravitational lensing study.  The study's initial sample combines 25 BCGs from the Cluster Lensing and Supernova Survey with Hubble (CLASH) sample and 37 BCGs from  the Sloan Giant Arcs Survey (SGAS) with a total redshift range of $0.2 < z < 0.7$  Using archival \textit{GALEX}, \textit{Hubble Space Telescope}, \textit{Wide-
Field Infrared Survey Explorer}, \textit{Herschel}, and Very Large Array data we determine the BCGs' stellar mass, radio power, and star formation rates. 
 The radio power is higher than expected if due to star formation, consistent with the BCGs being active galactic nucleus (AGN)-powered radio sources. 
This suggests that  the AGN and star formation are both fueled by cold gas in the  host galaxy. 
The specific star formation rate (sSFR) is low and constant with redshift. The mean sSFR is 9.42 $\times$ 10$^{-12}$ yr$^{-1}$, which corresponds to a mass doubling time of 105 billion years.
These findings are consistent with 
models for  hierarchical  formation of BCGs, which suggest that star formation is no longer a significant channel for galaxy growth for $z$ $\leq$ 1. Instead, stellar growth (of the order of a
factor of at least two) during this period is expected to occur mainly via minor dry mergers.
\end{abstract}

\section{Introduction}\label{sec:intro}
The most massive and luminous galaxies observed are the Brightest Cluster Galaxies (BCGs). They are typically located in the centers of galaxy clusters, indicating a relaxed position in the cluster potential. Their formation models are environmentally dependent, and distinct from typical elliptical galaxies \citep[e.g.,][]{Lin:2004aa,Brough:2005aa,De-Lucia:2007aa}.  BCGs are also more likely to host active galactic nuclei (AGNs) than other galaxies of the same stellar mass 
\citep[e.g.,][]{Best:2007aa,Von-Der-Linden:2007aa}. This indicates that these objects play a pivotal role in quenching cooling flows and star formation in clusters \citep{McNamara:2007aa,McNamara:2012aa,Fabian:2012aa}.
 BCGs have been recently shown to lie off the standard scaling relations of early-type galaxies 
\citep[e.g.,][]{Bernardi:2007aa,Lauer:2007aa,Von-Der-Linden:2007aa,Bernardi:2009aa}. In particular, they show excess luminosity (or stellar mass) above the prediction of the standard Faber--Jackson relation at high galaxy masses  \citep[e.g.,][]{Lauer:2014aa}. 

Stellar population synthesis models suggest  that the bulk of star formation in most massive galaxies took place prior to $z\sim  2$ \citep[e.g.,][]{Thomas:2005aa,Treu:2005aa,Jimenez:2007aa}. Semi-analytical models also suggest that the stars that make up most of the BCG mass are formed very early on (80\% by $z \sim 3$;  \citealt{De-Lucia:2006aa}, \citealt{De-Lucia:2007aa}).  Only after most stars have been formed does the final galaxy assembly take place via merging.  The final galaxy mergers in BCG formation ($z < 1$) are expected to be predominantly dissipationless (or dry, i.e., not involving large amounts of gas; \citealt{Khochfar:2003aa}, \citealt{De-Lucia:2007aa} ,\citealt{Vulcani:2016aa}).

Thus, a strong prediction of the hierarchical galaxy formation models is that at $z < 1$, star formation should make a only minor contribution to the growth of the stellar mass of BCGs.  This has been confirmed at low redshift \citep[e.g.,][]{Pipino:2009aa,Donahue:2010aa,Wang:2010aa,Liu:2012aa,Fraser-McKelvie:2014ab}. The vast majority of the BCGs studied previously have been at $z < 0.3$.  \citet{McDonald:2016aa} explored a much larger redshift range from $0.25 < z < 1.25$ and found that BCG star formation transitions from a disturbed cluster paradigm at $z > 0.6$ to a cool-core paradigm at $z = 0$.  They also find that at $z \sim 0.4$, 20\% of BCGs are forming stars at $> 10$ M$_{\odot}$ yr$^{-1}$.  To investigate this transition era, we present a study of star formation in two samples of BCGs (in clusters chosen for the study of gravitational lensing) that lie in the range $0.2 < z< 0.7$.

The outline for the paper is as follows.   In Section \ref{sec:sample}, we discuss the sample selection criteria. In Section \ref{sec:inst}, we discuss the archival  data used in the analysis. In Section \ref{sec:results}, we discuss how we estimated the star formation rates, stellar masses, and radio power for the BCGs.  In Section \ref{sec:disc}, we discuss the implications for formation models of BCGs. In  Section \ref{sec:conc}, we give our conclusions. 

This paper uses the current lambda cold dark matter ($\Lambda$CDM) parameters of  $H_{0}$ = 70 Mpc$^{-1}$ km s$^{-1}$, $\Omega_M$ = 0.3, and $\Omega_{vac}$ = 0.7.

\section{Sample Selection}\label{sec:sample}
We studied all  the BCGs from two samples of clusters selected for  studies of gravitational lensing - the Sloan Giant Arcs Survey \citep[SGAS:][]{Bayliss:2011aa} and the Cluster Lensing and Supernova Survey with Hubble \citep[CLASH:][]{Postman:2012aa}. The two samples cover a redshift range $0.2 < z < 0.7$. The clusters were not chosen on the basis of the star formation properties of the BCGs.  The coordinates,  redshifts, and modeled stellar mass of this BCG sample are described in Table 1 and their star formation rates from UV and optical methods are listed in Table 2.  

The 37 clusters in SGAS includes clusters selected from two different surveys and methods \citep{Oguri:2012ab}.  The first method is visual inspection of SDSS photometry of BCGs for strongly lensing systems (\citealt{Bayliss:2011aa}, M. D. Gladders et al., in preparation).  The second method is a ``blind study" in which SDSS $g$-band images of the 240 most massive clusters in the SDSS are selected based on lensing strength \citep{Hennawi:2008aa}. 

The selection criteria for the 25 CLASH clusters are discussed in detail by \citet{Postman:2012aa}. They selected 20 massive clusters from X-ray-based compilations of dynamically relaxed systems. Of these 20 clusters, 16 were taken from the \citet{Allen:2008aa}  compilation of massive relaxed clusters. An additional five clusters were added due to their exceptional strength as gravitational lenses (Einstein radii $>$ 35").
 The star formation properties of the CLASH BCGs are discussed independently \citep{Donahue:2015aa,Fogarty:2015aa} with results that are in agreement  with those reported here.

\section{Archival  Data}\label{sec:inst}
		\subsection{\textit{GALEX} Observations}
		In order to measure the population of young, massive stars in BCGs we obtained archival far-UV (FUV) and near-UV (NUV) band total pipeline photometric magnitudes from the GR6/GR7 release of the \textit{Galaxy Evolution Explorer} \citep[GALEX;][]{Martin:2005aa}.  GALEX photometry is taken using a 50 cm diameter Richey--Chretien telescope with a 1\fdg2 circular field of view.  To correct for galactic extinction, $E(B - V)$ values are retrieved from the GALEXview webpage for each of the target locations observed by GALEX \citep{Morrissey:2007aa}. Internal extinction is not corrected for due to H$\alpha$ being redshifted out of SDSS spectral coverage for much of our sample.  This prevents the use of the Balmer decrement to approximate internal dust levels across our entire sample.  Our sample K-corrections are not calculated in the UV due to the presence of limits in the SDSS observations necessary for k-correction calculation.  Previous CLASH UV findings \citep{Donahue:2015aa} have found k-corrections are less than 10\% and do not significantly effect results.  Archival data is a mixture of All Sky and Medium Imaging with depths of 20.5 and 23.5 $m_{AB}$ respectively.  For the farthest and least sensitive AIS case, MACS1149.6+2223, this corresponds to a SFR upper limit of 4.84 M$_{\odot}$ yr$^{-1}$.  Even in the worst-case scenario, this is still sensitive enough to detect examples of exceptionally high star formation and we only use these limits in cases with no optical emission line detections.  

The UV contribution from an old stellar population, called the ``UV-upturn" \citep{OConnell:1999aa}, can be addressed through modeling of the spectra of
other ellipticals in the BCG's cluster \citep[e.g.][]{Hicks:2010aa,Fogarty:2015aa}. In order to predict the old stellar UV contribution in a given target, we acquired a UV-optical-near-infrared (NIR) spectrum previously constructed using satellite ellipticals in the CLASH sample clusters \cite[][and K. Fogarty 2016, private communication]{Fogarty:2015aa}. These templates are composites of quiescent populations and do not include ongoing star formation components.  To approximate the UV from old stars,  \citet{Fogarty:2015aa} scaled their template to the available UV-Optical-NIR observations of the CLASH BCGs and derived a modeled J-band flux.  The correlation between J-band and NUV for old stellar populations in BCGs prescribed in Table 5 of \citet{Hicks:2010aa} was then used to generate an old stellar NUV flux.  We performed our SED scaling using the I-band from SDSS.  The Z-band was not used due to the more common occurrence of upper limits.   

		\subsection{Sloan Digital Sky Survey Observations (SDSS)}
		In order to measure the optical SED of the BCGs, we use the Catalog Archive Server Jobs System \citep[CASJOBS;][]{Li:2008aa} to obtain archival $ugriz$ Petrosian magnitudes \citep{Petrosian:1976aa} from the SDSS \citep{Gunn:2006aa}.  
  SDSS reports that the pipeline Petrosian magnitudes reliably include 80$\%$ of a galaxy's light independent of distance \citep{Blanton:2001aa}. We use Data Release 12 (DR12) results \citep{Alam:2015aa}. 	
 If the observed Petrosian magnitudes are lower than SDSS $ugriz$ limiting magnitudes [22.0, 22.2, 22.2, 21.3, 20.5], respectively, the galaxy is considered a non-detection and the Petrosian magnitude is used as a limit \citep{Ahn:2012aa}.  
We take extinction magnitudes from the photoObj SDSS tables to correct Milky Way foreground extinction.  Foreground extinction magnitudes assume a Milky Way Galaxy extinction map from \citep{Schlafly:2011aa} assuming a galactic reddening of $R_v$ = 3.1.  

	To measure the emission line features of the BCGs, we use archival spectral measurements and errors from SDSS DR12.  SDSS multi-object spectroscopy is taken using fibers that subtend 3" on the sky, large enough to include the majority of a given BCG.  The effects of this fixed aperture size, and its relation to the other apertures used for the photometry, are discussed in Section \ref{sec:results}.  Emission and absorption features in SDSS spectra are fit and stored in the galspecLine view in the publicly available Sky Server database of SDSS as a feature of the standard pipeline reduction of DR12. Foreground extinction was corrected for using the Milky Way extinctions calculated above.

		\subsection{\textit{WISE} Observations}
	The \textit{Wide-field Infrared Survey Explorer (WISE)} is a NASA space telescope \citep{Wright:2010aa} designed to perform an all sky survey using four bands.  To constrain the old stellar population during SED fitting, we retrieved \textit{WISE} profile fit 3.4, 4.6, 12 and 22 $\mu$m (W1, W2, W3, and W4) magnitudes from the NASA/IPAC Infrared Science Archive AllWISE Source Catalog \citep{Cutri:2013aa}.  The AllWISE catalog co-added exposures from the cryogenic and NEOWISE phases, resulting in 5$\sigma$ flux detections at 16.9, 16.0, 11.5, and 8.0 $m_{Vega}$ for W1, W2, W3, and W4, respectively.  Magnitudes are reported with photometric quality flags.  We treat data with flags U (95\% confidence upper limit) and C (2 $<$ SNR $<$ 3) as upper limits, and trust the stronger A (SNR $>$ 10) and B (3$<$ SNR $<$ 10) flags as detections.  Extinction in the infrared is negligible and is not corrected for in our work.  
		
		\subsection{Herschel Observations}
		To better characterize absorbed and re-emitted starlight from dust, we obtained archival data from the \textit{Herschel Space Observatory (Herschel)} --  a European Space Agency (ESA) built far infrared/sub-millimeter telescope launched in 2009 \citep{Pilbratt:2010aa}.  Archival Photoconductor Array Camera and Spectrometer (PACS) green band (100 $\mu$m), and red band (160 $\mu$m) photometry were acquired through the Herschel Science Archive.  A subset of CLASH galaxies were observed by the \textit{Herschel} Lensing Survey  \citep{Egami:2010aa} with flux limit of 2.4 mJy (100 $\mu$m) and 4.7 mJy (160 $\mu$m).  One SGAS target was observed by \citet{Rhoads:2014aa};  SDSSJ0915+3826 to limits of 5.2 mJy (100 $\mu$m) and 9.9 mJy (160 $\mu$m).  Another was observed by \citet{Saintonge:2013aa}; SDSSJ1343+4155 to limits of 4.42 mJy (70 $\mu$m) 9.9 mJy (160 $\mu$m).

To guarantee up-to-date calibrations for our faint sources, we re-reduced this sample's \textit{Herschel} PACS photometry with Herschel Interactive Programming Environment (HIPE) from the raw level 1 maps to the science quality level 2.5 using the MadMap pipeline \citep{Cantalupo:2010aa}.  This step is done in order to examine images with the latest \textit{Herschel} PACS calibration files.   Level 1 maps was retrieved from the Herschel Science Archive (HSA) with PACS calibration version PACS\_CAL\_69\_0.  MadMap removes the most common noise source in \textit{Herschel} PACS images, the 1/$f$ noise. 1/$f$ noise is the randomized photon noise over time the detector observes along with the target photons.  Once the level 2.5 maps are produced from HIPE's reduction process, the task {\tt{sourceExtractor}} was used to measure individual object fluxes and errors while correcting for sky background.  Afterward, we ran the {\tt sourceExtractor} task built into HIPE to detect objects with a S/N threshold of 3 using a gaussian with band-specific FWHMs of 10.5", 6.67", and 5.4" for red (160 $\mu$m), green (100 $\mu$m), and blue (70 $\mu$m) channels, respectively (PACS Handbook).  
	
		\subsection{Very Large Array (VLA) Observations}
	In order to quantify the activity of the supermassive black hole in each BCG, Karl G. Jansky VLA \citep{Thompson:1980aa} radio observations of observer-frame 1.4 GHz flux are taken from the 2014 version of the NRAO VLA Sky Survey \citep[NVSS;][]{Condon:1998aa} and the Faint Images of the Radio Sky at Twenty-Centimeters (FIRST) Survey Catalog \citep{Helfand:2015aa}.   NVSS is designed to observe 1.4 GHz emission at a low resolution of 45" across the entire sky north of $\delta$ = --40$^{\circ}$.  FIRST is a large-scale ($>$10,000 deg$^2$) survey that observed the radio sky at a resolution of 5" from 1993 to 2011 \citep{Helfand:2015aa}.  
	
	For objects with observations in both the NVSS and FIRST archival catalogs, preference is given to FIRST observations that have a better angular resolution as well as an rms of 0.15 mJy versus NVSS's rms of 0.45 mJy per beam.  Nine BCGs  observed by NVSS and FIRST do not have fully reduced photometry available in the published catalogs.  We used the Common Astronomy Software Applications  \citep{McMullin:2007aa} package to perform background subtraction and record the target radio flux or the upper limit if there is no visible source in the image.  Observations were k-corrected and extrapolated to target-frame 1.4 GHz using a spectral index of 0.75, common for BCGs \citep{Hlavacek-Larrondo:2013aa, Giacintucci:2014aa}.

\section{Analysis}\label{sec:results}
	BCG luminosity and stellar mass estimates were derived from the publicly available galaxy SED fitting software MAGPHYS \citep{da-Cunha:2008aa}.  MAGPHYS fits the SED of each galaxy using two components; UV to optical and NIR to FIR.  The fitting is done self-consistently, modeling the FIR flux considering the star formation fitted to the UV and optical.  Rather than the default \citet{Bruzual:2007aa} models, we use the \citet{Bruzual:2003aa} stellar population models found to successfully model BCGs in the past \citep[e.g.][]{Lidman:2012aa}. The infrared models include PAH, cold (15--25 K) , warm (30--60 K), and hot (130--250 K) dust components.  MAGPHYS fitted values used here include the total stellar mass.  Flux reduction due to cosmological expansion is accounted for by MAGPHYS before the fitting process using redshifts reported by \citet{Postman:2012aa} for CLASH targets and \citet{Alam:2015aa} for SDSS targets.  
	
	Five of the SGAS sample BCGs lacked SDSS photometric or spectroscopic redshifts at the time of this work and could not be accurately modeled: SDSSJ0333-0651, SDSSJ1420+3955, SDSSJ1522+2535, SDSSJ1621+0607, and SDSSJ2243-0935.  Additionally, SDSSJ0915+3826 and SDSSJ0928+2031 both had failed fits with infinite $\chi^2$.  This is believed to have occurred due to a combination of lack of NIR data as well as having the majority of its SDSS fluxes as limits.  These two are removed from the following statistics and plots.  An additional 13 targets had no SDSS spectra or GALEX photometry within 10", resulting in the final sample of 42 BCGs.
	
	Total star formation rate estimates from MAGPHYS are not used due to the poor infrared coverage of the SGAS sample.  The fitting procedure systematically overestimates the far infrared luminosity in the SGAS sample  due to lack of constraints.  This high estimated infrared luminosity corresponds to total star formation rates much higher than that which is consistent with the lack of dust and substructure seen in HST photometry.  However, the stellar mass estimates from MAGPHYS use the well-characterized optical-NIR continuum and are used further in this study.  For this study's results, we use the \citet{Kennicutt:1998aa} calibrations for optical emission line SFR estimates.  We use SDSS H$\alpha$ fluxes where possible, followed by [O II] when H$\alpha$ was redshifted out of the SDSS coverage.  GALEX NUV photometry was retrieved in the case where SDSS spectra are not available and we approximate the SFR using the \citet{Kennicutt:1998aa} UV continuum calibration.
	
	The SGAS subsample SFRs are predominantly approximated using H$\alpha$ and [O II] emission measured through the 3" diameter fibers of the SDSS multi-object spectrograph.  CLASH targets not observed by SDSS have SFRs approximated through GALEX NUV flux measured through elliptical Kron apertures in which the radius is 2.5 times the first moment of the target's radial profile.  The Kron radii used for targets in our sample range from 3" to 5" in radius.  CLASH SFRs are likely to be more accurate than SDSS fibers, which may miss star-forming knots in a given galaxy.  One recent example is \citet{Tremblay:2014aa} study of star formation filaments in a merging system of two massive ellipticals.  Slit spectroscopy using the ALFOSC spectrograph on the Nordic Optical Telescope found 1.7 times the SDSS H$\alpha$ flux reported in SDSS Data Release 9.  To correct for this effect would require a priori knowledge of the distribution, shape, and intensity of star-forming knots in all galaxies in our sample, so we leave the SDSS fluxes as reported by the pipeline.  Thus, the SGAS SFRs estimated using SDSS spectra will likely be lower limits to the true SFR.

	Additionally, as a sanity check for our MAGPHYS stellar mass results, we compare with previous estimates of the CLASH subsample from \citet{Burke:2015aa} and we are within 30\%.  For example, RXJ1532+30's stellar mass is estimated by \citet{Burke:2015aa} to be 2.20 $\pm$ 0.05 $\times$ 10$^{11}$ M$_{\odot}$ while we estimate 2.88 $\pm$ 0.06 $\times$ 10$^{11}$ M$_{\odot}$.  Their methodology uses magnitudes from the \citet{Postman:2012aa} observations fit to a spectral energy distribution produced by \citet{Bruzual:2003aa} solar metallicity models.   The combined CLASH-SGAS sample has a mean stellar mass of 7.52 $\times$ 10$^{11}$ M$_{\odot}$ and standard deviation of 6.22 $\times$ 10$^{11}$ M$_{\odot}$.

\section{Discussion}\label{sec:disc}

	\subsection{Dependencies With Mass}\label{sec:massdep}
\begin{figure}[h!]
  	\centering
	\includegraphics[width=0.45\textwidth]{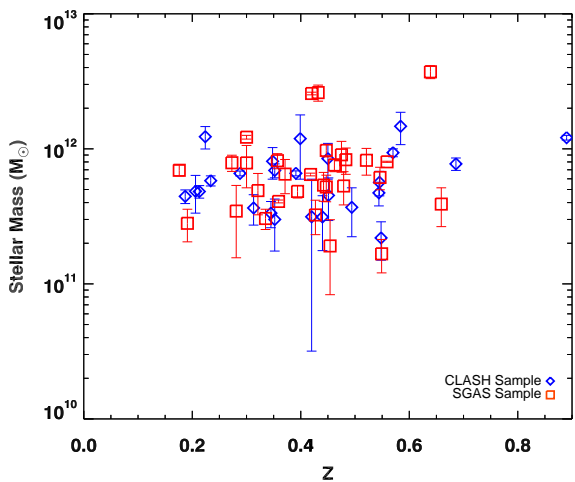}
	\caption{BCG stellar mass versus redshift of both samples as fit by MAGPHYS.  Red squares are SGAS sample galaxies.  Blue diamonds are CLASH sample galaxies.  Error bars are 1$\sigma$.}
	\label{fig:stellarmass}
\end{figure}	

\begin{figure}
	\centering
	\includegraphics[width=0.45\textwidth]{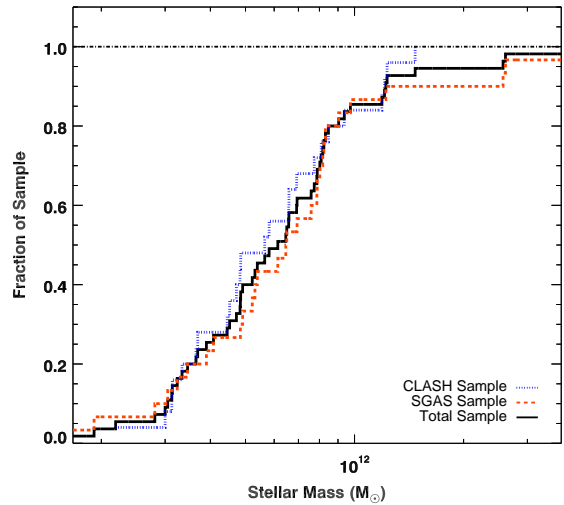}
	\caption{Cumulative histogram of stellar masses calculated for CLASH (dotted blue line), SGAS (dashed red line), and the combined sample (solid black line). }
	\label{fig:masshisto}
\end{figure}

  In Figure \ref{fig:stellarmass} we show the total stellar masses (estimated as described in \S~\ref{sec:results}) as a function of redshift. There is no trend of stellar mass with redshift  over the range $z$ = 0.2 - 0.7. We confirmed that the masses of the CLASH and SGAS samples are statistically similar using a  Kolmogorov--Smirnov test implemented using IDL's {\texttt {kstwo}} procedure on both samples. The probability the subsamples are identical is 0.75, with a maximum difference of 1.5\% between subsample distribution functions as seen in Figure \ref{fig:masshisto}.  
	
\begin{figure}[h!]
  	\centering
	\includegraphics[width=0.5\textwidth]{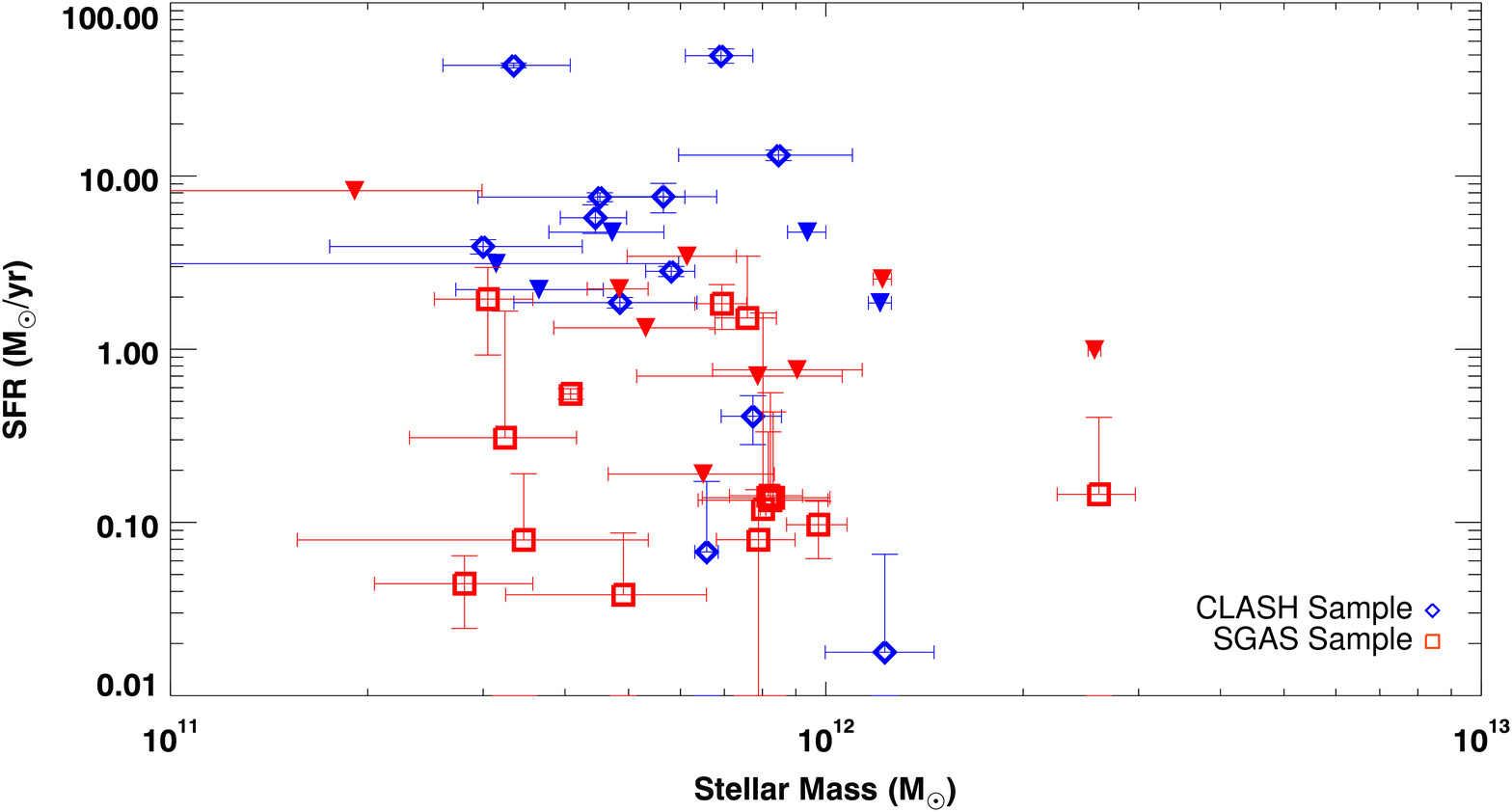}
	\caption{Star formation rate versus stellar mass of both samples as fit by MAGPHYS. Red squares are SGAS sample galaxies. Blue diamonds are CLASH sample galaxies. Error bars are 1$\sigma$.  Downward-facing triangles are upper limits. Missing lower limit bars are an artifact of plotting large errors on a log scale.}
	\label{fig:sfrvsmass}
\end{figure}

  In Figure \ref{fig:sfrvsmass} we examine the importance of star formation activity to a galaxy's evolution by comparing the star formation rate with the total stellar mass.  Previous results show that at redshifts $ 0 < z < 2$, star-forming galaxies in general show a correlation between stellar mass and star formation rate, which has been called a Star Formation Main Sequence \citep[SFMS; e.g.][]{Brinchmann:2004aa,Daddi:2007aa,Elbaz:2007aa,Noeske:2007aa, Santini:2009aa,Lara-Lopez:2010aa,Zahid:2013aa,Vaddi:2016aa}.     No SFMS trend is seen in our sample.  This corroborates previous results by \citet{Lee:2015aa} in which the SFMS is nearly flat (or absent) above 10$^{11}$ M$_{\odot}$.
	
		\subsection{Radio Behavior}\label{sec:radio}

\begin{figure}[h!]
	\centering
	\includegraphics[width=0.45\textwidth]{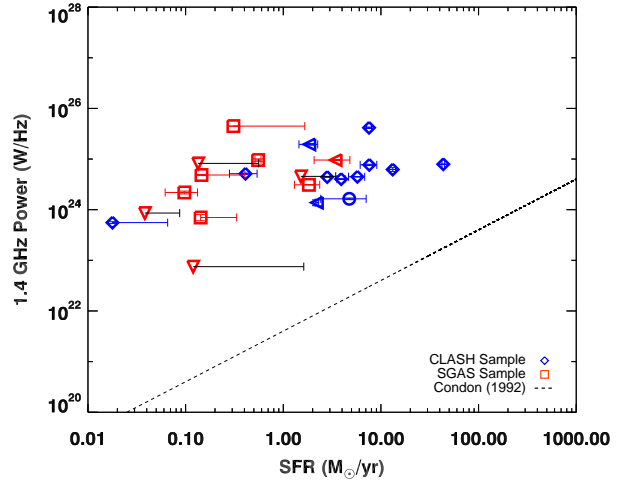}
	\caption{FIRST and NVSS 1.4 GHz rest-frame emission assuming $\alpha$ = 0.75.  The representative radio emission due to a star-formation-dominated scenario is plotted with the dashed line \citep{Condon:1992aa}.  Upper limits are the downward-facing triangles and from CASA-derived background-corrected photometry on target locations. Red squares are SGAS sample galaxies. Blue diamonds are CLASH sample galaxies.Error bars are 1$\sigma$. Downward-facing and leftward-facing triangles are upper limits for radio power and SFR, respectively.  Open circles indicate that both quantities are upper limits. Missing lower limit bars are an artifact of plotting large errors on a log scale.}
	\label{fig:radio}
\end{figure}

The observed  radio powers and upper limits of the BCGs (Figure \ref{fig:radio}) are one to two orders of magnitude above the radio powers expected if due solely to star formation estimated following \citet{Condon:1992aa}. This suggests that the radio emission is being powered by the AGNs in these BCGs. This is consistent with previous evidence that BCGs are often radio loud. The fraction of galaxies that are radio loud increases with galaxy mass \citep[e.g.,][]{Auriemma:1977aa,Dressel:1981aa,Ledlow:1996aa,Best:2005aa,Von-Der-Linden:2007aa}.   In addition, the average (and maximum) observed radio power increases with galaxy mass 
\citep[e.g.,][]{Brown:2011aa,Vaddi:2016aa}.
 This suggests that the source of fuel for AGN activity increases with galaxy mass; though radio power and mechanical power are not just a simple function of the accretion rate \citep[e.g.,][]{McNamara:2007aa,McNamara:2012aa,Best:2012aa}. 
Nevertheless, this is consistent with cold gas being the source of fuel for both the AGN and star formation activity. Cold gas has been suggested as the fuel for AGN activity 
\citep[e.g.,][]{Pizzolato:2005aa,Gaspari:2013aa,Gaspari:2015aa,Li:2014aa}.
There is evidence for infalling molecular gas in some BCGs, such as in A2597 \citep{Tremblay:2016aa}.

BCGs are more often found to be radio loud than non-BCG galaxies of the same mass \citep{Best:2007aa,Von-Der-Linden:2007aa}. This suggests that the BCGs have an additional source of gas that is not present in non-BCG galaxies of the same mass. Alternately, BCGs could have larger mass black holes than non-BCG galaxies of the same stellar mass \citep[e.g.,][]{Hlavacek-Larrondo:2012aa}.
The additional source of  gas in the BCGs is likely to be cooling from the hot ICM in cool-core clusters \citep[e.g.,][]{Fabian:1994aa,Best:2007aa,Sun:2009aa}.
The additional source of gas in the cool-core BCGs results in higher star formation than in the non-cool-core BCGs (see \S\ref{sec:z}).

		We find the range of BCG radio powers ($L_{1.4 GHz}$ $\approx$ 10$^{24-25}$ W Hz$^{-1}$) lie in the FR II  \citep{Fanaroff:1974aa,Bridle:1984aa} regime. These radio powers are sufficient to provide significant mechanical mode feedback to the environment, quenching star formation in the BCG 
\citep{McNamara:2007aa,McNamara:2012aa,Fabian:2012aa}.  
A radio power of 10$^{25}$ W Hz$^{-1}$ corresponds to a $\nu P_\nu$ Luminosity of 10$^{34}$ W. Assuming a 1\% efficiency of conversion of jet power to radio power \citep{Eilek:1989aa,Cavagnolo:2010aa,Daly:2012aa} and  10\% conversion of rest mass energy to jet power gives a mass accretion rate of $\sim 10^{-4}$ M$_\odot$ yr$^{-1}$, which is small compared to the SFR.

	\subsection{Dependencies with Redshift}\label{sec:z}
\begin{figure}
	\centering
	\includegraphics[width=0.5\textwidth]{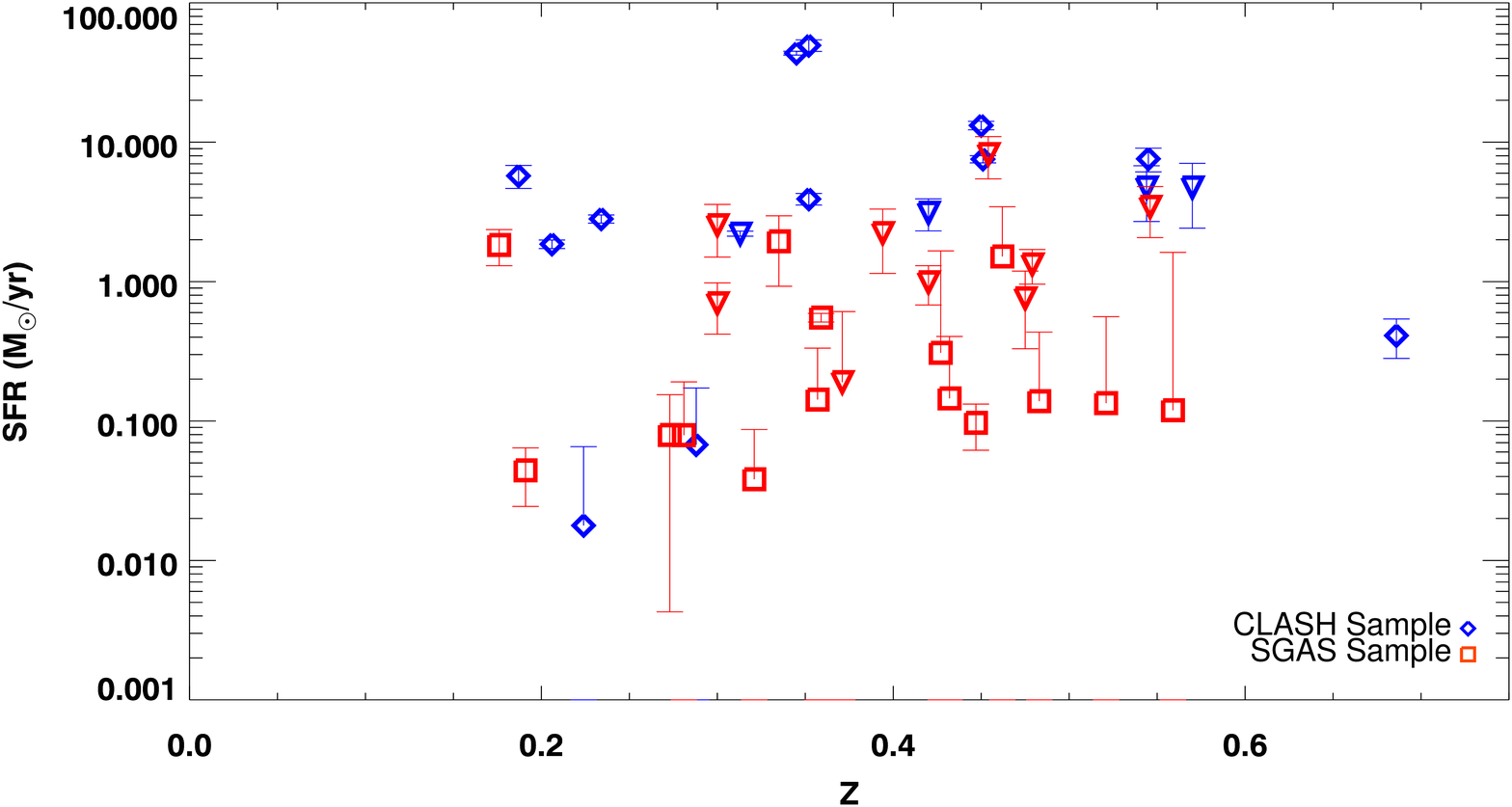}
	\caption{Star formation rate versus redshift of both samples.  Red squares are SGAS sample galaxies. Blue diamonds are CLASH sample galaxies.Error bars are 1$\sigma$. Downward-facing triangles are upper limits.  Missing lower limit bars are an artifact of plotting large errors on a log scale.}
	\label{fig:sfrvsz}
\end{figure}

	We plot star formation rates versus redshift in Figure \ref{fig:sfrvsz}. The estimated star formation rates show no trend with redshift over the range $z$ = 0.2 -- 0.6.  The SFR rates vary mostly between about 0.1 and 10  M$_{\odot}$ yr$^{-1}$. Examples of CLASH BCGs with high star formation rates are discussed in \citet{Fogarty:2015aa}. 
 
\begin{figure}[h!]
  	\centering
	\includegraphics[width=0.45\textwidth]{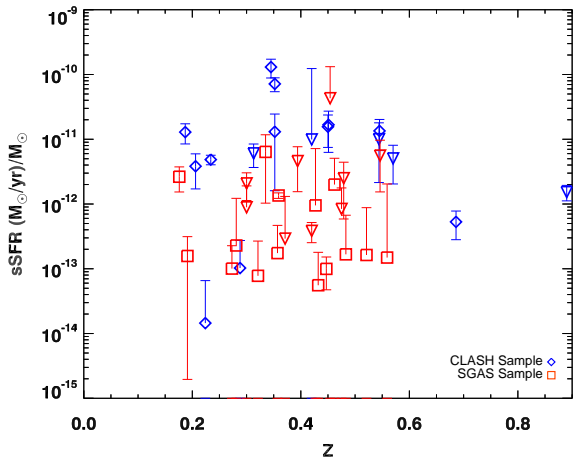}
	\caption{Specific star formation rate (sSFR) versus redshift of both samples.  Red squares are SGAS sample galaxies. Blue diamonds are CLASH sample galaxies.Error bars are 1$\sigma$. Downward-facing triangles are upper limits. Missing lower limit bars are an artifact of plotting large errors on a log scale.}
	\label{fig:specsfrvsz}
\end{figure}

We show the specific star formation rate (sSFR; star formation rate per unit stellar mass, e.g.,  Brinchmann et al. 2004) in  Figure \ref{fig:specsfrvsz}. There is no trend in sSFR with redshift over the range $z$ = 0.2 -- 0.6.
 There is slight difference in the sSFRs of the SGAS and CLASH samples.  For the combined MAGPHYS modeled data set of 17 CLASH BCGs and 24 SGAS BCGs, IDL's {\tt{twosampt}} task found that CLASH and SGAS have a probability of 49\% of being the same distribution.  This difference is likely due to the much smaller apertures used for measuring the H$\alpha$ flux used in the SFR determinations of the SGAS sample.

\begin{figure}[h!]
  	\centering
	\includegraphics[width=0.45\textwidth]{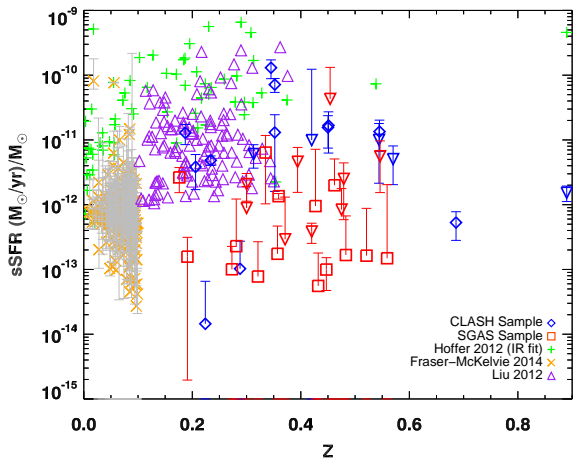}
	\caption{Specific star formation rate (sSFR) versus redshift, which includes published values for lower redshift samples.  Red squares are SGAS sample galaxies. Blue diamonds are CLASH sample galaxies.Error bars are 1$\sigma$. Downward-facing triangles are upper limits. Missing lower limit bars are an artifact of plotting large errors on a log scale.}
	\label{fig:litssfr}
\end{figure}

The mean sSFR is 9.42 $\times$ 10$^{-12}$ yr$^{-1}$ in our study, which corresponds to a mass doubling time of 105 billion years. Thus, the  low SFR and sSFR suggest that  galaxy growth via star formation in BCGs is minimal  by $z \sim 0.7$ consistent with other studies of star formation in lower redshift BCGs 
\citep[e.g.][]{Pipino:2009aa,Donahue:2010aa,Wang:2010aa,Liu:2012aa,Fraser-McKelvie:2014ab} as shown in Figure~\ref{fig:litssfr}.
 The outlier BCGs, which have higher than average star formation, tend to be in cool-core clusters
\citep[e.g.][]{Crawford:1999aa,Edwards:2007aa,Odea:2008aa,Odea:2010aa,Pipino:2009aa,Donahue:2010aa,Hicks:2010aa,Wang:2010aa, Pipino:2011aa,Liu:2012aa,Mittal:2015aa,Tremblay:2015aa,Donahue:2015aa,Fogarty:2015aa}.

\subsection{Growth by Mergers?}

\begin{figure}[h!]
  	\centering
	\includegraphics[width=0.45\textwidth]{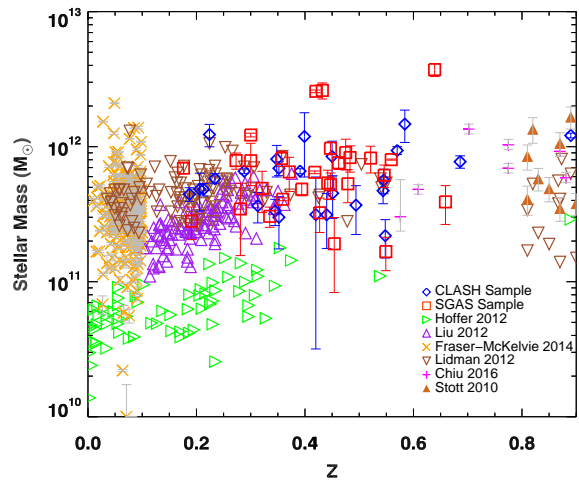}
	\caption{BCG stellar mass vs. redshift from different published samples of BCGs.  Red squares are SGAS sample galaxies. Blue diamonds are CLASH sample galaxies.  Error bars are 1$\sigma$. Downward-facing triangles are upper limits.  Missing lower limit bars are an artifact of plotting large errors on a log scale.}
	\label{fig:litmass}
\end{figure}

Growth of the mass of BCGs by mechanisms that do not involve star formation (e.g., dry mergers) is not  ruled out by this study of star formation. However, the lack of a redshift dependence on the BCG stellar mass (Figure \ref{fig:stellarmass}) rules out a \textit{large} gain in stellar mass over the redshift range $z$ = 0.3 -- 0.6 but probably allows a factor-of-two change in stellar mass over the redshift range $0 < z < 1$.  The previous observational work is somewhat inconsistent. Figure~\ref{fig:litmass} shows no clear trends in stellar mass with redshift, but the different samples have identified BCGs with different mass ranges.  \citet{Whiley:2008aa}  find no evidence for factor-of-two growth in BCG mass since $z \sim 1$. \citet{Collins:2009aa}  find no difference in mass between BCGs at $z \sim 1.3$ and $z \sim 0$. \citet{Stott:2011aa}  show that BCGs have evolved little in size since $z \sim 1$. \citet{Inagaki:2015aa}  find no evidence for more than a few percent mass growth between $z\sim 0.4$ and $z\sim 0.2$.
 In contrast, other work has found evidence for the expected increase in BCG stellar mass by about a factor of two between $z \sim 1$ and $z \sim 0$ 
\citep[e.g.,][]{Lidman:2012aa,Burke:2013aa,Lin:2013aa,Ascaso:2014aa,Shankar:2015aa}. Our specific sSFR results (Figure \ref{fig:litssfr}) corroborate \citet{McDonald:2016aa} results, which BCGs are evolving slower than other galaxies in cluster environments.  This would point to an alternate fuel source to keep BCGs mildly active when their satellites have been quenched.   The differences between the conclusions of these studies may be due to how the samples were defined and the effects of progenitor bias \citep[e.g.,][]{van-Dokkum:1996aa,Hopkins:2009aa,Saglia:2010aa,Carollo:2013aa,Shankar:2015aa}.

\begin{figure}[p]
	\centering
	\includegraphics[width=\textwidth]{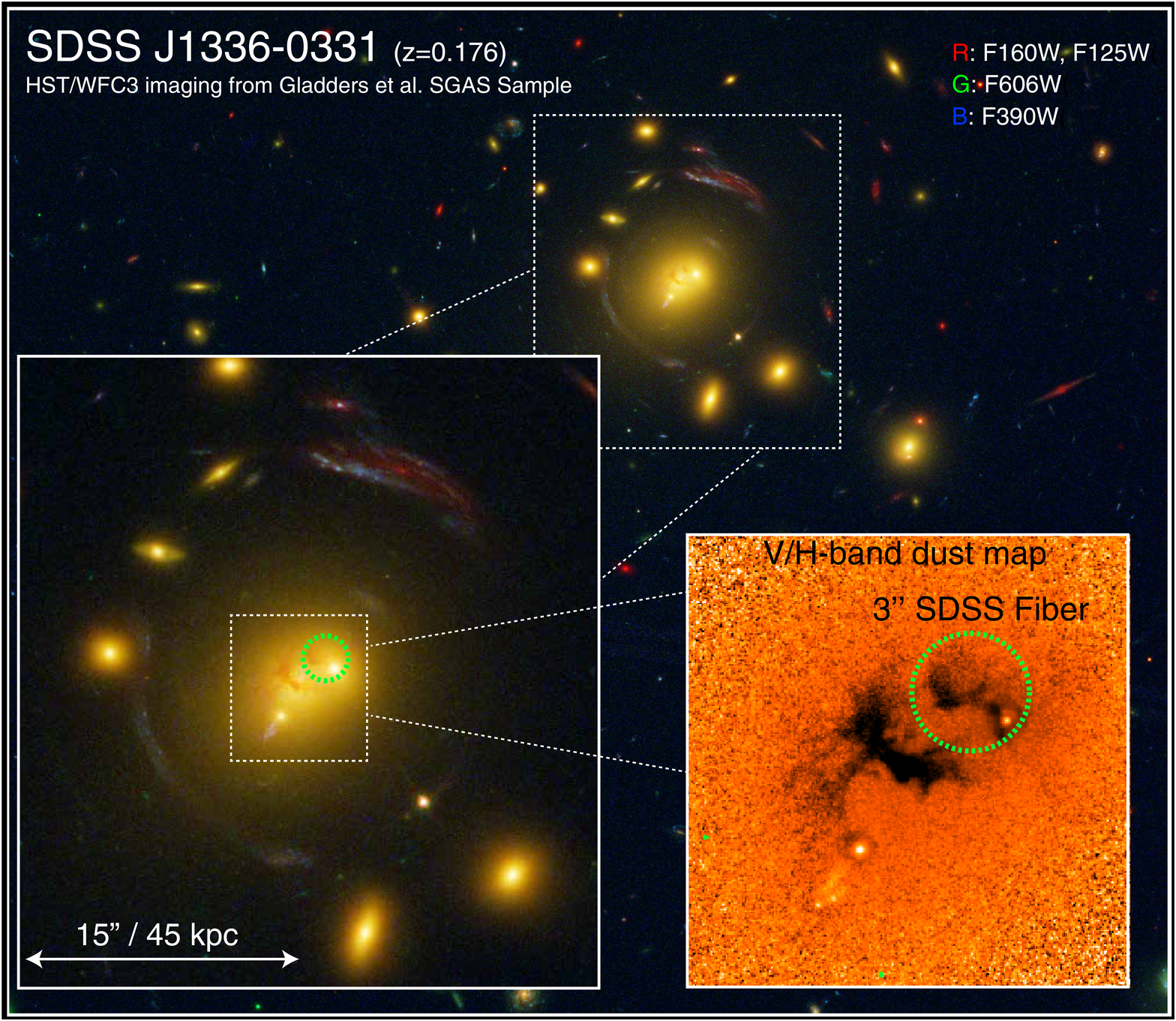}
	\caption{Left: \textit{Hubble Space Telescope} WFC3 three-color composite observation of cluster SDSSJ1336-0331 (PI:  Michael D. Gladders, ID: 13003) with an exposure time of 2400 s using filters F390W (blue), F606W (green), and F160W + F125W (red).  The green circle indicates the area subtended by the 3" diameter SDSS spectroscopic fiber.  Heavy dust obscuration covers the eastern nucleus, and a radial lensed feature is visible near the southeast.  The SDSS fiber misses any star formation in much of the southeast. 3" subtends 12.4 kpc at the target redshift.  Bottom right:  V/H-band color map revealing considerable substructure to the north and east along both nuclei.}
	\label{fig:1336}
\end{figure}

For the subset of star-forming BCGs of the CLASH subsample, the fuel for the star formation is attributed to gas condensing out of a cool core \citep{Donahue:2015aa,Fogarty:2015aa} which is consistent with the clusters having been X-ray selected. In the SGAS BCGs, X-ray evidence for cool cores is not yet available. We tentatively attribute the source of the fuel for star formation to a major merger in SDSSJ1336-0331 (Figure \ref{fig:1336}) and SDSSJ1531-3414 \citep{Tremblay:2014aa}. 

\section{Conclusions}\label{sec:conc}

We present multi-wavelength estimates of star formation rates, and estimates of radio power and stellar mass for BCGs in two samples of clusters chosen for the study of gravitational lensing - CLASH and SGAS. The redshift range of the BCGs $ 0.2< z < 0.7$ spans a large fraction of the range over which models of hierarchical galaxy formation suggest that star formation is no longer a significant channel for galaxy growth. Instead, stellar growth (of the order of a factor of at least two) during this period is expected to occur mainly via minor dry mergers \citep[e.g.,][]{De-Lucia:2007aa}.

We find that SFRs and sSFRs are indeed low in BCGs $0.2< z < 0.7$ (excluding some BCGs in cool-core clusters) consistent with results from samples of lower redshift BCGs 
\citep[e.g.,][]{Pipino:2009aa,Donahue:2010aa,Wang:2010aa,Liu:2012aa,Fraser-McKelvie:2014ab}. The mean sSFR is 9.42 $\times$ 10$^{-12}$ yr$^{-1}$, which corresponds to a mass doubling time of 105 billion years. This is in agreement with the results of the numerical and semi-analytical work 
\citep[e.g.,][]{Bell:2006aa,De-Lucia:2007aa} that galaxy growth by star formation is not significant at $z < 1$ in BCGs.

Based on their high radio power, the BCGs are AGN-powered radio sources, consistent with expectations for BCGs \citep{Best:2007aa,Von-Der-Linden:2007aa}. This  is consistent with the AGN and star formation in these sources  both  being  fueled by cold gas in the  host galaxy.

\acknowledgements{
\linespread{1}

We thank the referee for constructive comments which helped make this paper a more robust study.  We would like to thank the referee for constructive comments which helped make this paper a more robust study.  We thank Kevin Fogarty of Johns Hopkins University for his advice on the nature of old stars in the UV and his UV model of SFR-poor elliptical satellites. We thank Elisabete Da Cunha for her upper-limit-capable MAGPHYS code as well as her advice in its implementation. GRT acknowledges support from the NASA through Einstein Postdoctoral Fellowship Award Number PF-150128, issued by the Chandra X-ray Observatory Center, which is operated by the Smithsonian Astrophysical Observatory for and on behalf of NASA under contract NAS8-03060.  Funding for SDSS-III has been provided by the Alfred P. Sloan Foundation, the Participating Institutions, the National Science Foundation, and the U.S. Department of Energy Office of Science. The SDSS-III web site is http://www.sdss3.org/. SDSS-III is managed by the Astrophysical Research Consortium for the Participating Institutions of the SDSS-III Collaboration including the University of Arizona, the Brazilian Participation Group, Brookhaven National Laboratory, Carnegie Mellon University, University of Florida, the French Participation Group, the German Participation Group, Harvard University, the Instituto de Astrofisica de Canarias, the Michigan State/Notre Dame/JINA Participation Group, Johns Hopkins University, Lawrence Berkeley National Laboratory, Max Planck Institute for Astrophysics, Max Planck Institute for Extraterrestrial Physics, New Mexico State University, New York University, Ohio State University, Pennsylvania State University, University of Portsmouth, Princeton University, the Spanish Participation Group, University of Tokyo, University of Utah, Vanderbilt University, University of Virginia, University of Washington, and Yale University. Based on observations made with the NASA \textit{Galaxy Evolution Explorer}. 
$GALEX$ is operated for NASA by the California Institute of Technology under NASA contract NAS5-98034. Based on observations made with the NASA/ESA \textit{Hubble Space Telescope}, obtained from the Data Archive at the Space Telescope Science Institute, which is operated by the Association of Universities for Research in Astronomy, Inc., under NASA contract NAS 5-26555. These observations are associated with program \#13003. This publication makes use of data products from the \textit{Wide-field Infrared Survey Explorer}, which is a joint project of the University of California, Los Angeles, and the Jet Propulsion Laboratory/California Institute of Technology, funded by the National Aeronautics and Space Administration. \textit{Herschel} is an ESA space observatory with science instruments provided by European-led Principal Investigator consortia and with important participation from NASA. HCSS / HSpot / HIPE is a joint development (are joint developments) by the Herschel Science Ground Segment Consortium, consisting of ESA, the NASA Herschel Science Center, and the HIFI, PACS, and SPIRE consortia. The National Radio Astronomy Observatory is a facility of the National Science Foundation operated under cooperative agreement by Associated Universities, Inc.}


\newcommand\invisiblesection[1]{%
  \refstepcounter{section}%
  \addcontentsline{toc}{section}{\protect\numberline{\thesection}#1}%
  \sectionmark{#1}}

\invisiblesection{Bibliography}

\bibliography{ResearchbibdeskFeb16}

\newpage

\invisiblesection{Tables}

\begin{deluxetable}{lccccccr}


\tabletypesize{\footnotesize}


\tablecaption{CLASH and SGAS BCG Samples}

\tablenum{1}

\tablehead{\colhead{Target ID} & \colhead{$\alpha_{J2000}$} & \colhead{$\delta_{J2000}$} & \colhead{Redshift} & \colhead{Stellar Mass} &  \colhead{$\chi^2$} & \colhead{Sample} \\ 
\colhead{} & \colhead{(degrees)} & \colhead{(degrees)} & \colhead{($z$)}  & \colhead{($M_{\odot}$)} & \colhead{} & \colhead{}} 
\label{table:coordinates}
\startdata
Abell 209 & 01:31:52.6 & -13:36:38.80 & 0.206 & 4.85e+11 & 0.599 & CLASH \\
Abell 383 & 02:48:03.4 & -03:31:44.7 & 0.187 & 4.45e+11 & 5.331 & CLASH \\
Abell 611 & 08:00:56.8 & +36:03:24.1 & 0.288 & 6.58e+11 & 146.998 & CLASH \\
Abell 1423 & 11:57:17.3 & +33:36:37.4 & 0.213 & 4.83e+11 & 3.308 & CLASH \\
Abell 2261 & 17:22:27.2 & +32:07:58.6 & 0.224 & 1.23e+12 & 12.845 & CLASH \\
CLJ1226.9+3332 & 12:26:58.4 & +33:32:47.4 & 0.89 & 1.21e+12 & 5.748 & CLASH \\
MACS0329.7-0211 & 03:29:41.7 & -02:11:47.7 & 0.45 & 8.47e+11 & 0.235 & CLASH \\
MACS0416.1-2403 & 04:16:09.4 & -24:04:03.9 & 0.42 & 3.14e+11 & 0.452 & CLASH \\
MACS0429.6-0253 & 04:29:36.1 & -02:53:08.0 & 0.399 & 1.19e+12 & 0.004 & CLASH \\
MACS0647.8+7015 & 06:47:50.0 & +70:14:49.7 & 0.584 & 1.47e+12 & 0.054 & CLASH \\
MACS0710.5+3745 & 07:17:31.7 & +37:45:18.5 & 0.548 & 2.19e+11 & 0.015 & CLASH \\
MACS0744.9+3927 & 07:44:52.8 & +39:27:24.4 & 0.686 & 7.74e+11 & 1.175 & CLASH \\
MACS1115.9+0129 & 11:15:52.1 & +01:29:56.6 & 0.352 & 3e+11 & 2.1 & CLASH \\
MACS1149.6+2223 & 11:49:35.9 & +22:23:55.0 & 0.544 & 4.72e+11 & 1.907 & CLASH \\
MACS1206.2-0847 & 12:06:12.3 & -08:48:02.4 & 0.44 & 3.13e+11 & 1.129 & CLASH \\
MACS1211.0-0210 & 13:11:01.7 & -03:10:39.5 & 0.494 & 3.69e+11 & 1.068 & CLASH \\
MACS1423.8+2404 & 14:23:47.8 & +24:04:40.5 & 0.545 & 5.65e+11 & 1.145 & CLASH \\
MACS1720.3+3536 & 17:20:17.0 & +35:36:23.6 & 0.391 & 6.59e+11 & 11.689 & CLASH \\
MACS1931.8-2635 & 19:31:49.7 & -26:34:34.0 & 0.352 & 6.92e+11 & 12.782 & CLASH \\
MACS2129.4-0741 & 21:29:26.1 & -07:41:28.8 & 0.57 & 9.37e+11 & 8.04 & CLASH \\
MS2137-2353 & 21:40:15.2 & -23:39:40.7 & 0.313 & 3.65e+11 & 0.242 & CLASH \\
RXJ1347.5-1145 & 13:47:30.6 & -11:45:10.1 & 0.451 & 4.52e+11 & 0.026 & CLASH \\
RXJ1532.9+3021 & 15:32:53.8 & +30:20:58.7 & 0.345 & 3.34e+11 & 12.048 & CLASH \\
RXJ2129.7+0005 & 21:29:39.9 & +00:05:18.8 & 0.234 & 5.81e+11 & 3.317 & CLASH \\
RXJ2248.7-4431 & 22:48:44.3 & -44:31:48.4 & 0.348 & 8.09e+11 & 0.065 & CLASH \\
SDSSJ0004-0103 & 00:04:51.9 & -01:03:15.80 & 0.479 & 5.31e+11 & 0.941 & SGAS \\
SDSSJ0108+0624 & 01:08:42.0 & +06:24:43.50 & 0.549 & 1.67e+11 & 4.877 & SGAS \\
SDSSJ0146-0929 & 01:46:56.0 & -09:29:52.40 & 0.447 & 9.74e+11 & 1.036 & SGAS \\
SDSSJ0150+2725 & 01:50:00.9 & +27:25:36.20 & 0.3 & 1.22e+12 & 171.751 & SGAS \\
SDSSJ0851+3331 & 08:51:39.0 & +33:31:10.83 & 0.371 & 6.5e+11 & 2.036 & SGAS \\
SDSSJ0915+3826 & 09:15:39.0 & +38:26:58.77 & 0.396 & ... & 999.0 & SGAS \\
SDSSJ0928+2031 & 09:28:05.6 & +20:31:25.55 & 0.192 & ... & 999.0 & SGAS \\
SDSSJ0952+3434 & 09:52:40.0 & +34:34:47.09 & 0.359 & 4.08e+11 & 3.055 & SGAS \\
SDSSJ0957+0509 & 09:57:39.2 & +05:09:31.80 & 0.447 & 5.22e+11 & 0.283 & SGAS \\
SDSSJ1002+2031 & 10:02:26.9 & +20:31:02.61 & 0.321 & 4.91e+11 & 0.358 & SGAS \\
SDSSJ1038+4849 & 10:38:43.2 & +48:49:18.73 & 0.432 & 2.61e+12 & 8.098 & SGAS \\
SDSSJ1050+0017 & 10:50:39.9 & +00:17:06.91 & 0.3 & 7.87e+11 & 3.303 & SGAS \\
SDSSJ1055+5547 & 10:55:04.6 & +55:48:23.23 & 0.462 & 7.59e+11 & 2.304 & SGAS \\
SDSSJ1110+6459 & 11:10:17.6 & +64:59:47.02 & 0.659 & 3.9e+11 & 1.592 & SGAS \\
SDSSJ1115+1645 & 11:15:04.4 & +16:45:38.40 & 0.191 & 2.81e+11 & 2.971 & SGAS \\
SDSSJ1138+2754 & 11:38:09.0 & +27:54:30.90 & 0.454 & 1.91e+11 & 0.221 & SGAS \\
SDSSJ1152+0930 & 11:52:47.3 & +09:30:14.54 & 0.521 & 8.23e+11 & 0.537 & SGAS \\
SDSSJ1152+3313 & 11:52:00.3 & +33:13:41.72 & 0.357 & 8.17e+11 & 2.322 & SGAS \\
SDSSJ1156+1911 & 11:56:05.5 & +19:11:12.68 & 0.546 & 6.14e+11 & 0.314 & SGAS \\
SDSSJ1207+5254 & 12:07:36.4 & +52:54:58.20 & 0.273 & 7.89e+11 & 9.278 & SGAS \\
SDSSJ1209+2640 & 12:09:23.7 & +26:40:46.50 & 0.559 & 8.02e+11 & 2.811 & SGAS \\
SDSSJ1329+2243 & 13:29:34.5 & +22:43:00.24 & 0.42 & 2.57e+12 & 89.54 & SGAS \\
SDSSJ1336-0331 & 13:36:00.0 & -03:31:28.63 & 0.176 & 6.94e+11 & 53.666 & SGAS \\
SDSSJ1343+4155 & 13:43:32.8 & +41:55:04.48 & 0.418 & 6.46e+11 & 41.59 & SGAS \\
SDSSJ1439+1208 & 14:39:09.9 & +12:08:24.75 & 0.427 & 3.24e+11 & 0.735 & SGAS \\
SDSSJ1456+5702 & 14:56:00.8 & +57:02:20.60 & 0.483 & 8.31e+11 & 0.672 & SGAS \\
SDSSJ1527+0652 & 15:27:45.4 & +06:52:31.79 & 0.394 & 4.84e+11 & 0.6 & SGAS \\
SDSSJ1531+3414 & 15:31:10.6 & +34:14:24.91 & 0.335 & 3.05e+11 & 0.477 & SGAS \\
SDSSJ1604+2244 & 16:04:10.2 & +22:44:16.69 & 0.281 & 3.46e+11 & 1.368 & SGAS \\
SDSSJ1632+3500 & 16:32:10.3 & +35:00:30.16 & 0.475 & 9.04e+11 & 0.511 & SGAS \\
SDSSJ1723+3411 & 17:23:36.2 & +34:11:59.37 & 0.442 & 5.39e+11 & 0.487 & SGAS \\
SDSSJ2111-0114 & 21:11:19.4 & -01:14:23.57 & 0.639 & 3.72e+12 & 12.035 & SGAS \\
\enddata

\tablerefs{\citet{da-Cunha:2008aa}, \citet{Postman:2012aa},
M.D. Gladders et al. in preparation}
\tablecomments{Columns:  Target, R.A.(J2000), Decl.(J2000),  SDSS redshift, MAGPHYS estimated  stellar mass ($M_{\odot}$), fit $\chi^2$, and original subsample. $\ast$ denotes that RXJ2248.7-4431 is alternately named Abell 1063S. }

\end{deluxetable}

\begin{deluxetable}{lccccccr}


\tabletypesize{\footnotesize}


\tablecaption{CLASH and SGAS BCG Star Formation Rate Estimators}

\tablenum{2}

\tablehead{\colhead{Target ID} & \colhead{NUV SFR} & \colhead{NUV SFR Error} & \colhead{NUV Quality} & \colhead{H$\alpha$ SFR} & \colhead{H$\alpha$ SFR Error} &  \colhead{[O II] SFR} &  \colhead{[O II] SFR Error} \\ 
\colhead{} & \colhead{($M_{\odot}$yr$^{-1}$)} & \colhead{($M_{\odot}$yr$^{-1}$)} & \colhead{}  & \colhead{($M_{\odot}$yr$^{-1}$)} & \colhead{($M_{\odot}$yr$^{-1}$)} & \colhead{($M_{\odot}$yr$^{-1}$)} & \colhead{($M_{\odot}$yr$^{-1}$)}} 
\label{table:coordinates}
\startdata
Abell 209 & 1.85 & 0.13 & D & ... & ... & ... & ... \\
Abell 383 & 5.74 & 1.08 & D & ... & ... & ... & ... \\
Abell 611 & ... & ... & N & ... & ... & 0.06 & 0.10 \\
Abell 1423 & ... & ... & N & ... & ... & ... & ... \\
Abell 2261 & ... & ... & N & 0.09 & 0.07 & 0.01 & 0.04 \\
CLJ1226.9+3332 & 1.85 & 0.402 & L & ... & ... & ... & ... \\
MACS0329.7-0211 & 13.20 & 0.92 & D & ... & ... & ... & ... \\
MACS0416.1-2403 & 3.11 & 0.80 & L & ... & ... & ... & ... \\
MACS0429.6-0253 & ... & ... & N & ... & ... & ... & ... \\
MACS0647.8+7015 & ... & ... & N & ... & ... & ... & ... \\
MACS0710.5+3745 & ... & ... & N & ... & ... & ... & ... \\
MACS0744.9+3927 & ... & ... & N & 0.41 & 0.12 & ... & ... \\
MACS1115.9+0129 & 5.38 & 0.59 & D & 2.83 & 0.19 & 3.91 & 0.37 \\
MACS1149.6+2223 & 4.74 & 2.04 & L & ... & ... & ... & ... \\
MACS1206.2-0847 & ... & ... & N & ... & ... & ... & ... \\
MACS1211.0-0210 & ... & ... & N & ... & ... & ... & ... \\
MACS1423.8+2404 & 7.6 & 1.47 & D & ... & ... & ... & ... \\
MACS1720.3+3536 & ... & ... & N & ... & ... & ... & ... \\
MACS1931.8-2635 & 49.51 & 4.72 & D & ... & ... & ... & ... \\
MACS2129.4-0741 & 4.74 & 2.32 & L & ... & ... & ... & ... \\
MS2137-2353 & 2.21 & 0.09 & L & ... & ... & ... & ... \\
RXJ1347.5-1145 & 7.55 & 0.44 & D & ... & ... & ... & ... \\
RXJ1532.9+3021 & ... & ... & N & 31.88 & 0.38 & 43.53 & 1.36 \\
RXJ2129.7+0005 & 1.71 & 0.64 & D & 1.71 & 0.05 & 2.81 & 0.19 \\
RXJ2248.7-4431 & ... & ... & N & ... & ... & ... & ... \\
SDSSJ0004-0103 & 1.33 & 0.37 & L & ... & ... & ... & ... \\
SDSSJ0108+0624 & ... & ... & N & ... & ... & ... & ... \\
SDSSJ0146-0929 & ... & ... & N & 0.09 & 0.03 & ... & ... \\
SDSSJ0150+2725 & 2.54 & 1.04 & L & ... & ... & ... & ... \\
SDSSJ0851+3331 & 0.19 & 0.42 & L & ... & ... & ... & ... \\
SDSSJ0915+3826 & 1.91 & 0.55 & L & 0.28 & 0.127 & 0.03 & 1.02 \\
SDSSJ0928+2031 & 0.17 & 0.06 & L & 0.16 & 0.08 & ... & ... \\
SDSSJ0952+3434 & 1.72 & 0.67 & L & 0.55 & 0.039 & ... & ... \\
SDSSJ0957+0509 & ... & ... & N & ... & ... & ... & ... \\
SDSSJ1002+2031 & ... & ... & N & 0.03 & 0.04 & ... & ... \\
SDSSJ1038+4849 & 7.6 & 2.04 & L & ... & ... & 0.14 & 0.25 \\
SDSSJ1050+0017 & 0.7 & 0.28 & L & ... & ... & ... & ... \\
SDSSJ1055+5547 & ... & ... & N & ... & ... & 1.51 & 1.92 \\
SDSSJ1110+6459 & ... & ... & N & ... & ... & ... & ... \\
SDSSJ1115+1645 & 0.25 & 0.18 & L & 0.04 & 0.01 & 0.04 & 0.17 \\
SDSSJ1138+2754 & 8.22 & 2.76 & L & ... & ... & ... & ... \\
SDSSJ1152+0930 & ... & ... & N & ... & ... & 0.13 & 0.42 \\
SDSSJ1152+3313 & ... & ... & N & 0.10 & 0.10 & 0.14 & 0.19 \\
SDSSJ1156+1911 & 3.44 & 1.37 & L & ... & ... & ... & ... \\
SDSSJ1207+5254 & 0.28 & 0.13 & L & 0.20 & 0.10 & 0.07 & 0.07 \\
SDSSJ1209+2640 & ... & ... & N & ... & ... & 0.11 & 1.50 \\
SDSSJ1329+2243 & 0.99 & 0.31 & L & ... & ... & ... & ... \\
SDSSJ1336-0331 & 3.76 & 1.473 & D & ... & ... & 1.83 & 0.53 \\
SDSSJ1343+4155 & ... & ... & N & ... & ... & ... & ... \\
SDSSJ1439+1208 & ... & ... & N & ... & ... & 0.30 & 1.35 \\
SDSSJ1456+5702 & ... & ... & N & ... & ... & 0.13 & 0.29 \\
SDSSJ1527+0652 & 2.23 & 1.085 & L & ... & ... & ... & ... \\
SDSSJ1531+3414 & ... & ... & N & 1.60 & 0.44 & 1.94 & 1.01 \\
SDSSJ1604+2244 & ... & ... & N & ... & ... & 0.07 & 0.11 \\
SDSSJ1632+3500 & 0.76 & 0.43 & L & ... & ... & ... & ... \\
SDSSJ1723+3411 & ... & ... & N & ... & ... & ... & ... \\
SDSSJ2111-0114 & ... & ... & N & ... & ... & ... & ... \\
\enddata

\tablecomments{Columns:  Target ID, NUV star formation rate and error; NUV data quality (N - no data, L - 95\% Confidence Upper Limit, D-detection); H$\alpha$ SFR and error; and [O II] SFR and error.  All star formation rates are in units of M$_{\odot}$yr and errors represent 1$\sigma$ values.}

\end{deluxetable}

\end{document}